\documentclass[twocolumn,prl,amsmath,amssymb,showpacs,superscriptaddress]{revtex4-1}
\usepackage{epsf}      
\usepackage{graphicx}
\usepackage{color}
\usepackage{soul}
\usepackage{gensymb}

\begin{document}

\title {Electrochemical properties of Na$_{0.66}$V$_4$O$_{10}$ nanostructures as cathode material in rechargeable batteries for energy storage applications}

\author{Rakesh Saroha}
\affiliation{Department of Physics, Indian Institute of Technology Delhi, Hauz Khas, New Delhi-110016, India}
\author{Tuhin S. Khan}
\affiliation{Department of Chemical Engineering, Indian Institute of Technology Delhi, Hauz Khas, New Delhi-110016, India}
\author{Mahesh Chandra}
\affiliation{Department of Physics, Indian Institute of Technology Delhi, Hauz Khas, New Delhi-110016, India}
\author{Rishabh Shukla}
\affiliation{Department of Physics, Indian Institute of Technology Delhi, Hauz Khas, New Delhi-110016, India}
\author{Amrish K. Panwar}
\affiliation{Department of Applied Physics, Delhi Technological University, Delhi-110042, India}
\author{Amit Gupta}
\affiliation{Department of Mechanical Engineering, Indian Institute of Technology Delhi, Hauz Khas, New Delhi-110016, India}
\author{M. Ali Haider}
\affiliation{Department of Chemical Engineering, Indian Institute of Technology Delhi, Hauz Khas, New Delhi-110016, India}
\author{Suddhasatwa Basu}
\affiliation{Department of Chemical Engineering, Indian Institute of Technology Delhi, Hauz Khas, New Delhi-110016, India}
\author{Rajendra S. Dhaka}
\email{rsdhaka@physics.iitd.ac.in}
\affiliation{Department of Physics, Indian Institute of Technology Delhi, Hauz Khas, New Delhi-110016, India}

\date{\today}                                         

\begin{abstract}

{\bf ABSTRACT:} We report the electrochemical performance of nanostructures of Na$_{0.66}$V$_4$O$_{10}$ as cathode material for rechargeable batteries. The Rietveld refinement of room temperature x-ray diffraction pattern shows the monoclinic phase with C2/m space group. The cyclic voltammetry curves of prepared half-cells exhibit redox peaks at ~3.1 and 2.6~V, which are due to two-phase transition reaction between V$^{5+/4+}$ and can be assigned to the single step deintercalation/intercalation of Na-ion. We observe a good cycling stability with specific discharge capacity (measured vs. Na$^+$/Na) between 80 ($\pm$2) and 30 ($\pm$2) mAh g$^{-1}$ at a current density 3 and 50~mA g$^{-1}$, respectively. The electrochemical performance of Na$_{0.66}$V$_4$O$_{10}$ electrode was also tested with Li anode, which showed higher capacity, but decay faster than Na. Using density functional theory, we calculate the Na vacancy formation energies; 3.37~eV in the bulk of the material and 2.52~eV on the (100) surface, which underlines the importance of nanostructures.     

\end{abstract}

\maketitle

\section{\noindent ~Introduction}
After commercialization in early 1990s, rechargeable Li-ion batteries (LIBs) are being widely investigated and used as energy storage to power portable electronic devices and hybrid electric vehicles \cite{Goodenough13,ReddyCR13}. This is mainly because LIBs show the lowest redox potential of the Li (E$_{\rm Li/Li^+}$ = -3.04 V vs. SHE), which allows possessing high voltage and energy density. Also, small ionic radius of Li$^+ $($\sim$0.76 \AA~in six coordination state) allow its smooth diffusion during charging/discharging, which makes long cycle life \cite{Han14,MaromJMC11}. However, due to huge demand, high cost, safety concerns, limited and non-uniform Li resources \cite{GoodenoughCM10}, the Na-ion batteries (NIBs) have generated considerable interest as the most promising alternative (due to the uniform distribution and abundance of Na resources in the earth crust) to the LIBs for large-scale energy storage systems \cite{KimAEM12, PanEES13, SlaterAFM13, KimAEM16, LarcherNC15,VaalmaNRM18,HwangCSR17}. The elements Na and Li are from the same group in the periodic table and have similar properties such as reactivity, physical strength, etc. The global abundance, low cost and appropriate redox potential (E$_{\rm Na/Na^+}$ = -2.71 V vs. SHE) prove its suitability as a good substitute for Li. However, large ionic radius of Na-ion ($\sim$1.02 \AA~in the six coordination environment) as compared to the Li leads to slow ionic diffusion and lower energy density \cite{SawickiRSC15, NayakAC18}. Owing to these facts, materials with open framework or materials which can exchange more than one Na per formula unit, or Na-rich materials are crucial for the possibility of having high power and energy density \cite{KunduAC15,ChenASS18}. 

In this context, layered transition metal oxides having formula NaTO$_2$ (T= transition metal) and polyanionic compounds have been extensively explored and used as a cathode material for Na-ion battery \cite{ToumarPRA15, YuPRA17, ZhengPRA17, DebbichiPRB15, RushPRM17, YabuuchiNM12, XiangAM15, NiASN17}. For example, vanadium oxychloride \cite{GaoEC15}, Na$_{0.44}$Mn$_{1-x}$Zn$_x$O$_2$ \cite{CaoAM11,MaheshCI18}, NaNi$_{1/3}$Mn$_{1/3}$Co$_{1/3}$O$_2$ \cite{SathiyaCM12}, Na$_3$V$_2$(PO$_4$)$_3$ \cite{LuluJPS14}, olivine-type phospahte and sulphate based cathodes like NaFePO$_4$ \cite{OhEC12}, NaFeSO$_4$F \cite{BarpandaIC10} and even prussian blue derivatives which contain a suitable transition metal such as KTFe(CN)$_6$ \cite{PalomaresEES13}, which acts as an open host framework with a large interstitial space to absorb bigger Na$^+$. However, due to their complex reaction mechanism and in order to improve thermal stability, energy density and cycling performance, it is vital to search new cathode materials and investigate their physical/electro-chemical properties for Na-ion batteries \cite {ZhuNS13, Mahesh18}.  

Interestingly, in recent years, vanadium based oxide materials have attracted worldwide attention as a possible alternate cathode material for both Li-ion and Na-ion batteries \cite{NiJPS14, NiJMCA15, NiAMI16, YuanJAC16, LiuJPS11, Lee17, Wang16}.  Because of the variable oxidation state of vanadium (+5 in V$_{2}$O$_{5}$ to +2 in VO), it acts as electron donor and acceptor during the process of Na extraction and insertion \cite{PalomaresJMCA15}. The crystal structure of V$_2$O$_5$ consists of three different vanadium sites labeled as V(1), V(2) and V(3) \cite{LiuJPS11}. The edge sharing vanadium octahedra [V(1)O6] and corner sharing vanadium octahedral [V(2)O6] combine to form zigzag and double chains along the $b$ axis, respectively. The edge sharing [V(3)O5] pyramid forms connection to the [V$_4$O$_{11}$] in layers, which are formed from the oxygen linkage to the V(1) and V(2) octehedra, thus yielding a 3D tunnelled structure \cite{LuACS15}. The formed 3D tunnelled structure is not only more stable as compared to the layered structures, but also provide fast ion reaction kinetics. It has been reported that depending upon the Li$^+$/Na$^+ $ intercalation amount, V$_2$O$_5$ can yield high discharge capacities \cite{HuRSCA17,PanJMC10}. In Li-ion batteries, Li$_x$V$_2$O$_5$ is explored extensively \cite{JiangJPCC07, WangEA15, AslEA13, SemenenkEC10}, but it suffers significant capacity loss during repetitive charging/discharging for an intercalated amount of $x >$1 \cite{PanJMC10}.  Analogous to Li$_x$V$_2$O$_5$ in Li-ion,  Na$_x$V$_2$O$_5$ was also proposed for Na-ion battery, where the Na cations can be reversibly cycled along the $b$ axis in a similar manner as the Li counterpart. Within pentoxide based cathode, {Na$_{0.33}$V$_2$O$_5$ has gained significant attention as a cathode material for Li-ion/Na-ion batteries because of high experimental discharge capacity \cite{BachSSI89,KimACS15, LuACS15, LiangEA14}. Bach, {\it et. al.} have first investigated the Na intercalation and de-intercalation behavior in Na$_{0.33}$V$_2$O$_5$ bronze \cite{BachSSI89} and then similarly, NaV$_6$O$_{15}$ [(Na$_{0.33}$V$_2$O$_5$)$_3$] have been studied as a cathode material in various morphologies such as nanoplates, nanoroads, nanoflowers, and nanoflakes for Li/Na-ion batteries by various research groups \cite{WangRSC16, LiuJPS11, HuRSCA17, JiangJES15}. Jiang et al. synthesized {NaV$_6$O$_{15}$ nanoplates and reported a discharge capacity of 116~mAh g$^{-1}$ measured vs. Na$^+$/Na with a cycle retention of 55\% at a current density of 50~mA g$^{-1}$ \cite{JiangJES15}. However, note that the electrochemical performance of Na$_{0.33}$V$_2$O$_5$ critically depends on the synthesis conditions, size in nanometer and morphology \cite{WangRSC16, LiuJPS11, HuRSCA17, JiangJES15}. Therefore, nanostructures of Na$_{0.66}$V$_4$O$_{10}$ [(Na$_{0.33}$V$_2$O$_5$)$_2$] in different morphologies still need to be explored in detail as cathode material for Na-ion batteries. Also, it have been reported that partial substitution of Li with Na in Li$_{2-x}$Na$_x$MnO$_3$ improve the electrochemical performance and significantly enhance the cycling stability of Li-ion battery \cite{DongJPC13}. This further motivates to test our Na$_{0.66}$V$_4$O$_{10}$ cathode with Li as counter electrode.

In this paper, we synthesize nanorods of Na$_{0.66}$V$_4$O$_{10}$ and study the electrochemical performance for a promising cathode material in Na-ion batteries. The Rietveld refinement of x-ray diffraction pattern reveals the monoclinic phase with C2/m space group. The transmission electron microscopy (TEM) and scanning electron microscopy (SEM) studies reveal the rod shaped morphology with agglomeration. The size and the length of the rods were found to be 50-100~nm and few $\mu$m, respectively. The cyclic voltammetry (CV) curves of prepared half-cells exhibit redox peaks at $\approx$ 3.1 and 2.6~V, which indicate phase transition reaction in V$^{5+/4+}$ and can be assigned to the single step deintercalation/intercalation of Na-ion. Interestingly, we observed good cycling stability with specific discharge capacity of 80 ($\pm$2) and 30 ($\pm$2) mAh g$^{-1}$ at a current density of 3 and 50~mA g$^{-1}$, respectively, measured vs. Na$^+$/Na anode. We have also tested the electrochemical performance with Li anode. The energetics of Na vacancy formation on discharging and Na vacancy filling on charging could form a potential descriptor for electrochemical performance. Since Na vacancy formation was relatively easier on the electrode surface, a surface diffusion pathway for Na transport could be hypothesized. For nanostructures, the available surface area for this pathway is likely to increase, which further alludes to the measured electrochemical performance. 

\section{\noindent ~Results and Discussion}

\subsection{\noindent ~Structural Analysis}

In order to decide the annealing temperature for stable phase formation, we performed the thermal analysis of the resulting gel precursor under air atmosphere using thermogravimetric analysis (TGA) profile. Figure~\ref {fig:TGA} shows the variation in mass (\%) of the gel precursor with increase in temperature. The mass loss of around 10\% below 180$\degree$C (region~1) is due to the removal of physically adsorbed as well as some intercalated water molecules \cite{ManikandanEA16}. The steep mass loss ($\Delta \rm m=$ 46.3\%) observed between 180-400$\degree$C in region~2 is attributed to the removal of acetate precursors and volatile impurities in the form of NH$_3$ and CO$_2$ gases. A vigorous oxidation and decomposition reactions take place in this region leading to the phase formation of NVO material. Further, negligible mass loss was observed for region 3 and 4, i.e. above 400$\degree$C in TGA profile. It can be observed from TGA profile that annealing at about 400$\degree$C would be sufficient for the phase formation of Na$_{0.66}$V$_4$O$_{10}$ (NVO) material. We have also tried annealing at higher temperatures and found good thermal stability of the prepared material.

\begin{figure}
\includegraphics[width=3.3in]{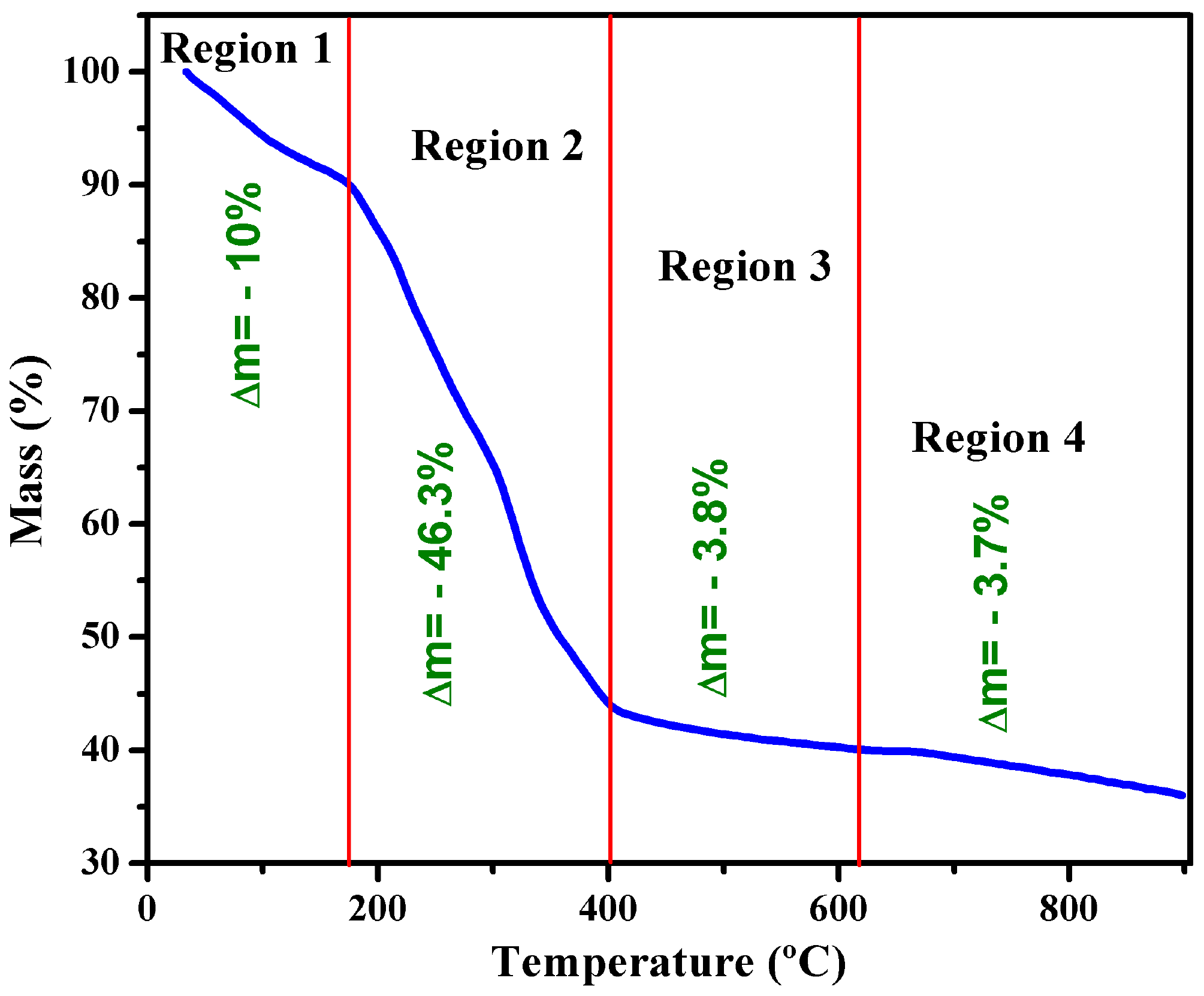}
\caption{Thermogravimetric analysis (TGA) of the precursors used to synthesize Na$_{0.66}$V$_4$O$_{10}$ samples.}
\label{fig:TGA}
\end{figure}

\begin{figure}
\includegraphics[width=3.6in]{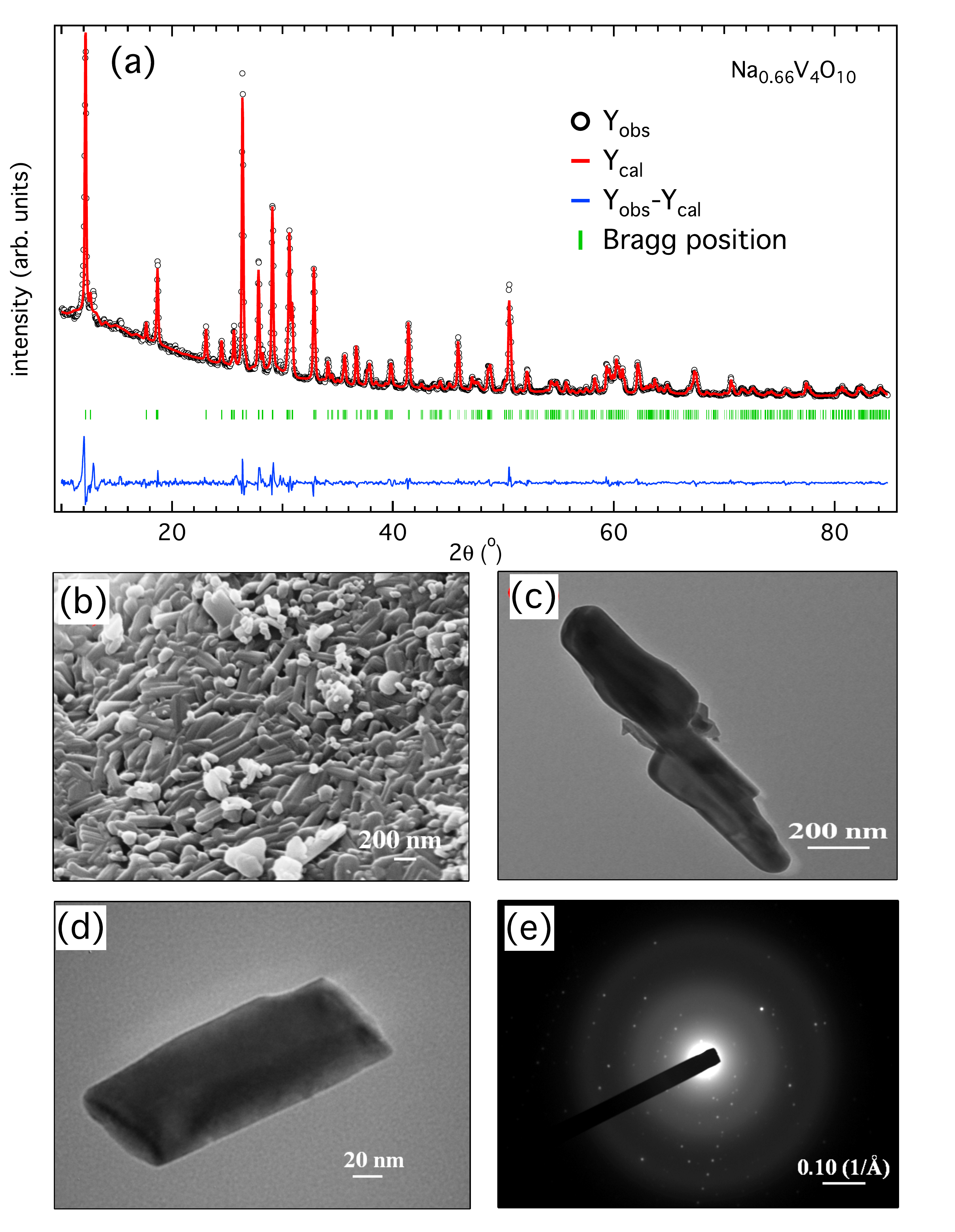}
\caption{Structural and morphological properties of prepared Na$_{0.66}$V$_4$O$_{10}$ electrode material: (a) XRD pattern with Rietveld refinement, (b) SEM micrograph, (c, d) TEM images at different scale, and (e) SAED pattern. The label 0.10 (1/$\rm \AA$) is the scale bar in reciprocal lattice is equivalent to 1 (1/nm).} 
\label{fig:XRDF}
\end{figure} 

The x-ray diffraction (XRD) pattern along with the corresponding Rietveld refinement of the prepared Na$_{0.66}$V$_4$O$_{10}$ (NVO) sample are shown in Figure~2(a). The XRD pattern was matched with the standard phase using X-pert high score plus software with PDF2 reference data file. The obtained XRD pattern of the synthesized NVO powder could be well-indexed with the pure phase of the pristine sodium vanadium bronze i.e. $\beta$-Na$_{0.33}$V$_2$O$_5$ having monoclinic structure with C2/m space group (JCPDS code: 48-0382), as also reported in ref.~\cite{GrzechnikJPCM16}. The sharp and intense peaks reveal the crystalline nature of the sample. The Rietveld refinement was performed to obtain the quantitative crystal structure information such as lattice parameters. The peaks were modelled using the Pseudo-Voigt function and the full width at half maximum (FWHM) was refined using Lorentzian broadening of the XRD peaks. The structural refinement results ($\chi$$^2$=2.47, R$_p=$ 4.83\%, R$_{wp}=$ 6.6\%) indicate that the fitting is reliable and good. The obtained lattice parameters from refinement are: $a$ = 15.426~\AA, $b$ = 3.610~\AA, $c$ = 10.073~\AA~ and $\beta$= 109.55$^o$ are in good agreement with previous reports of Na$_{0.33}$V$_2$O$_5$ \cite{BaddourJMC11}. Further, we have provided the fractional coordinates and site occupancies in the following Table~I.

\begin{table}[h]
		\centering
		\label{tab:rietveld}
		\caption{Atomic coordinates, site occupancies and isotropic displacement parameters for as-prepared Na$_{0.66}$V$_4$O$_{10}$, obtained from the Rietveld refinement of room temperature x-ray diffraction spectrum.}
		\begin{tabular}{|c|c|c|c|c|c|c|}
		\hline
		\text{atom}&\text{site}&\text{x}&\text{y}&\text{z}&\text{B$_{\rm iso} (\rm\AA^2$})&Occ. \\
		\hline
 Na & 4\textit{i} &0.0045(12) &0.00000 & 0.4100(16) & 0.5 & 0.775\\
 V1 & 4\textit{i} & 0.3382(3) & 0.00000 & 0.0987(5) & 0.5 & 1.337\\
 V2 & 4\textit{i} & 0.1196(3) & 0.00000 & 0.1198(5) & 0.5 & 1.334\\ 
 V3 & 4\textit{i} & 0.2885(3) & 0.00000 & 0.4059(6) & 0.5 & 1.334\\
 O1 & 2\textit{a} & 0.00000 & 0.00000 & 0.00000 & 0.5 & 0.977\\
 O2 & 4\textit{i} &0.8156(10) & 0.00000 & 0.0509(17) & 0.5 & 1.202\\
 O3 & 4\textit{i} &0.6345(9) & 0.00000 & 0.0796(16)  & 0.5 & 1.295\\
 O4 & 4\textit{i} &0.4334(9) & 0.00000 & 0.2192(15)  & 0.5 & 1.394\\
 O5 & 4\textit{i} &0.2666(9) & 0.00000 & 0.2175(17) & 0.5 & 1.299\\
 O6 & 4\textit{i} &0.1078(9) & 0.00000 & 0.2599(15)& 0.5 & 1.357\\
 O7 & 4\textit{i} &0.2391(11) & 0.00000 & 0.5709(16)& 0.5 & 1.229\\
 O8 & 4\textit{i} & 0.3981(9) & 0.00000 & 0.4597(16)& 0.5 & 1.254\\
		\hline
		\end{tabular}
\end{table}

\begin{figure*}
\includegraphics[width=7.3in]{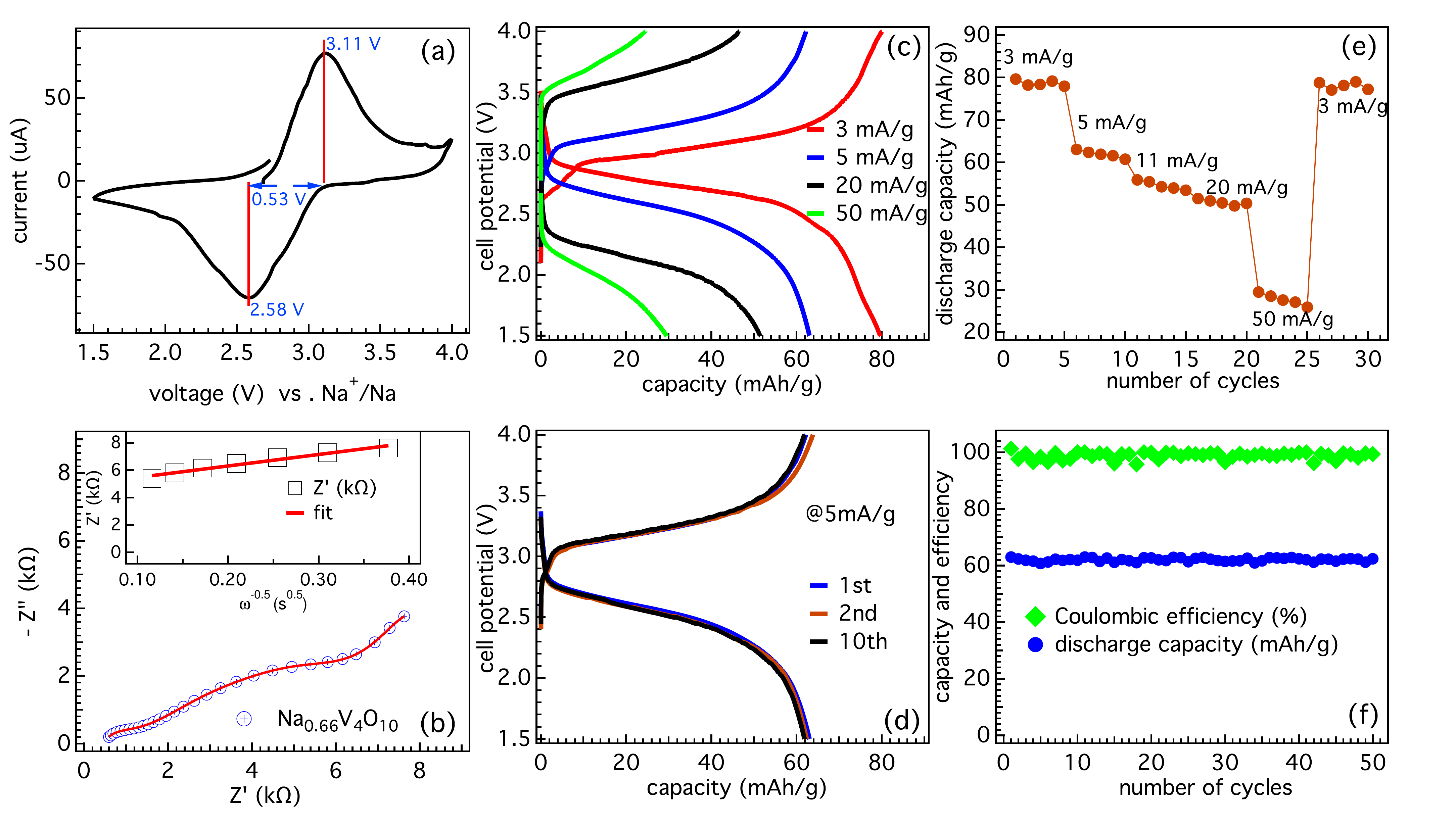}
\caption{Electrochemical properties of prepared NVO cathode with Na anode: (a) Cyclic voltammetry (CV) at a scan rate of 0.05 mV s$^{-1}$ vs. Na$^+$/Na at room temperature, (b) Electrochemical impedance spectra of as prepared NVO sample at AC amplitude of 5 mV, inset shows a plot between Z' and $\omega^{-0.5}$. (c) Initial charge/discharge curves at different current densities, (d) 1st, 2nd and 10th charge/discharge curve at 5~mA g$^{-1}$ current density, (e) The rate performance at different current densities, (f) The cycling performance and Coulombic efficiency for 50 cycles measured at 5~mA g$^{-1}$.} 
\label{fig:Electrochemical}
\end{figure*}

We further characterize the prepared NVO cathode material to check the morphological properties. The SEM micrograph reveals the formation of uniformly distributed highly agglomerated rod-shaped morphology, as shown in Figure~2(b). However, the length of the rods varies dramatically, which give rise to some plate-like particles over rod-shaped morphology. The TEM images in Figures~2(c, d) clearly confirm the formation of nano-sized rod-shaped morphology of the NVO sample, in agreement with the SEM analysis. The dark and slightly greyish regions correspond to the NVO nanorods, which display uniform dimensions with 50--100~nm wide and several micrometers long. The cathode materials in the nanosized particles/rods shape are expected to enhance the electrochemical performance of Na-ion batteries. Note that the characteristic diffusion length (L) of nanostructures and the diffusion coefficient (D) of Na-ion are related as: $\tau$ = L$^2$/D, where $\tau$ is the time for Na-ion to diffuse through the electrode material. In this context, one can achieve better electrochemical performance by decreasing L, which effectively reduce $\tau$. The selected area electron diffraction pattern (SAED) for the synthesized NVO sample is shown in Figure~2(e), which reveals the existence of sharp diffraction spots along with diffused circles which represents polycrystllaine nature of the synthesized sample. The interplanar spacing/distance ($d$ = 0.47~nm) calculated using SAED pattern matches to the (200) peak of the XRD pattern.     

\subsection{\noindent ~Electrochemical Performance}

Now we present, in Figures~3(a--d), the electrochemical performance of the prepared NVO cathode in terms of cyclic voltammetry, galvanostatic charge/discharge curves, rate performance and Coulombic efficiency. Figure~3(a) shows the cyclic voltammograms (CV) of prepared NVO electrode in the potential range of 1.5--4.0~V at a scan rate of 0.05~mV s$^{-1}$, measured v/s Na$^+$/Na at room temperature. Interestingly, it exhibits redox peaks at around 3.1 and 2.6~V, which are due to two-phase transition reactions between V$^{5+/4+}$ and can be assigned to the single step de-intercalation/intercalation of Na-ion in the Na$_{0.66}$V$_4$O$_{10}$ electrode. It is interesting to note that no such characteristics have been observed in vanadium based oxide cathodes. Liu {\it et al.} reported the charge/discharge profile of NaV$_6$O$_{15}$ [(Na$_{0.33}$V$_2$O$_5$)$_3$], however, that have multiple steps/plateaus \cite{LiuJPS11}. The obtained CV results are consistent with charge-discharge profiles, as shown in Figure~3(c). The insertion and extraction behaviour of Na-ion can be tentatively given as:
\begin{equation}
Na_{0.66}V_4O_{10} \longleftrightarrow Na_{0.66-x}V_4O_{10} + xNa^+ + xe^-
\end{equation}

Figure~3(c) shows the galvanostastic first charge-discharge curves of prepared CR2016 half-cells using NVO as cathode and recorded at various current densities from 3~mA g$^{-1}$ to 50~mA g$^{-1}$ in the potential range of 1.5--4.0~V vs. Na$^+$/Na. The specific discharge capacity of the NVO sample is observed to be 80 ($\pm$2), 64 ($\pm$2), 56 ($\pm$2), 52 ($\pm$2) and 30 ($\pm$2) mAh g$^{-1}$ at a current density of 3, 5, 11, 20 and 50~mA g$^{-1}$, respectively. Assuming no electrolyte decomposition during the first charge, Na$_{0.66}$V$_4$O$_{10}$ charged to Na$_{0.09}$V$_4$O$_{10}$ (corresponding to the charge capacity of 80~mAh g$^{-1}$ at 3~mA g$^{-1}$) and then, back to Na$_{0.66}$V$_4$O$_{10}$. Note that the discharge capacity of about 80~mAh g$^{-1}$ (1C equated to 141~mAh g$^{-1}$) corresponds to +4.86 oxidation state of vanadium, consistent with the CV results. It can also be seen from the Figure~3(c) that as current density increases from 3~mA g$^{-1}$ to 50~mA g$^{-1}$, the specific discharge capacity value decreases from 80 ($\pm$2) to 30 ($\pm$2) mAh g$^{-1}$ with an apparent drop in potential plateau, which is due to the increase in polarization and resistance or iR drop of the electrode at high current densities. Note that the OCV for NVO cathode was observed to be around 2.8~V. The GCD results in Fig.~3(c) show the charge/discharge voltage plateau around 2.9/2.8~V, respectively at 3~mA g$^{-1}$ vs. Na/Na$^+$. The relatively high over potential indicates that the activation energy of the redox event is very high because of which the potential necessary to transfer an electron from NVO cathode is relatively high even at low current density. The comparison of 1st, 2nd and 10th cycle discharge capacity curves are shown in Figure~3(d). The discharge capacity remains fairly constant for the ten cycles, which elaborates the structural stability of the electrode material. The cycling stability and rate performance for an electrode are very important parameters to determine the stability in long-term applications. Figsure~3(e, f) show the stepwise electrochemical rate and cycling performance, respectively. As observed from Figure~3(e), the capacity decreases with increase in current density and goes to minimum for 50~mA g$^{-1}$. Figure~3(f) shows the cycling performance of the synthesized NVO sample at 5~mA g$^{-1}$ current density at room temperature. It shows an initial discharge capacity 64 ($\pm$2) mAh g$^{-1}$ and retains nearly 100\% capacity even after 50 charge/discharge cycles. The Coulombic efficiency of the NVO electrode for 50 charge/discharge cycles at 5~mA g$^{-1}$ current density is also shown in Figure~3(f). The Coulombic efficiency for all the charge/discharge cycles is observed to be more than 95\% [Figure~3(f)]. 

Furthermore, the electrochemical impedance spectra (EIS) of fresh cells were performed to measure the electrode/electrolyte resistance. Figure~3(b) shows the impedance spectra of NVO sample measured with an AC voltage pulse having amplitude of 5~mV. The EIS curve, measured in the frequency range of 10~mHz to 100~kHz, shows a depressed semi-circle and a straight line in the high and low-frequency regions, respectively. In general, the ohmic resistance (R$_s$) of the cell, i.e. the resistance due to electrolyte and electrode material, can be estimated by an intercept at the Z$^{'}$-axis in the high-frequency region. The information about the electrochemical reactions taking place at the electrode/electrolyte interface, which indicate the charge transfer resistance (R$_{ct}$), can be obtained from the semi-circle in the middle-frequency range. Also, the Warburg impedance (Z$_w$), which is associated with Na-ion diffusion in the electrode active material, can be represented by the inclined line in the low-frequency region. The value of cell parameters were obtained to be R$_s$= 594~$\ohm$ and R$_{ct}$=  778~$\ohm$. 

The chemical diffusion coefficient of the Na ions inside an electrode material is calculated using the following equation: 
\begin {equation}
D=\frac{T^2R^2}{2n^4A^2F^4C^2\sigma^2_w}
\end{equation} 
where, D, T, R, $n$, A, C, and F are diffusion coefficient (cm$^2$ s$^{-1}$), absolute temperature (K), gas constant (8.314 J mol$^{-1}$ K$^{-1}$), number of electrons involved in the redox process, electrode area (2.2~cm$^2$), the Na-ion concentration (3.139$\times$10$^{-3}$ mol cm$^{-3}$), and the Faraday constant (96486 C mol$^{-1}$), respectively. The $\sigma_w$ is the Warburg impedance coefficient, which is related to Z$^{'}$ by the following equation:
\begin {equation}
Z'=R_s+R_{ct}+\sigma_w \omega^{-0.5}
\end{equation} 
In the inset of Figure~3(b), we show the obtained value of $\sigma_w$ from the slope of Z$^{'}$ and $\omega^{-0.5}$, which found to be 7739 $\Omega$ s$^{-0.5}$. The value of $n$ (=59~mV/$\Delta V$) can be calculated by taking the difference between oxidation and reduction peaks ($\Delta V=$ 0.53 V) in CV [see Figure~3(a)], which found to be 0.1 (number of electrons participating in charging and discharging process). Finally by using these values in equation~2, the obtained value of the diffusion coefficient is 1.239$\times$10$^{-13}$ cm$^2$ s$^{-1}$, which is in good agreement with the one reported in ref.~\cite{HuCI18}.

It is important to compare the observed capacity in this manuscript with the other related cathode materials reported in the literature \cite{Liunanoscale14,ZhangJPS18,ShindeJPS19}. For example, Liu {\it et al.}, observed the discharge capacity of about 100~mAhg$^{-1}$ at 0.1~C current rate in Na$_3$V$_2$(PO$_4$)$_3$/C nanofibers cathode materials \cite{Liunanoscale14}. Zhang {\it et al.}, found the reversible capacity of about 120~mAhg$^{-1}$ for NaVO$_3$ cathode after the activation at high-voltage \cite{ZhangJPS18}. More recently, Shinde {\it et al.}, used the ultrasonic sonochemical synthesis method to prepare Na$_{0.44}$MnO$_2$ cathode, which shows the reversible capacity of about 110~mAhg$^{-1}$ at a current rate of C/10 with good cycling stability \cite{ShindeJPS19}. By changing the preparation method to self-combustion, the NaV$_6$O$_{15}$ nanoplates show the capacity of about 150~mAhg$^{-1}$ at a current rate of 20~mAg$^{-1}$, which remains about 82~mAhg$^{-1}$ after 30 cycles \cite{JiangJES15}. On the other hand, NaV$_6$O$_{15}$ nanorods prepared by PVP-modulated synthesis route shows high initial capacity of about 157~mAhg$^{-1}$ at 20~mAg$^{-1}$ current density \cite{Wang16}. It should be noted that the discharge capacity of NVO cathode crucially depend on the preparation method, amount of Na content in the material and morphology of the nanoparticles as well as type of intercalation Li$^+$/Na$^+$.

\begin{figure*}
\includegraphics[width=7.3in]{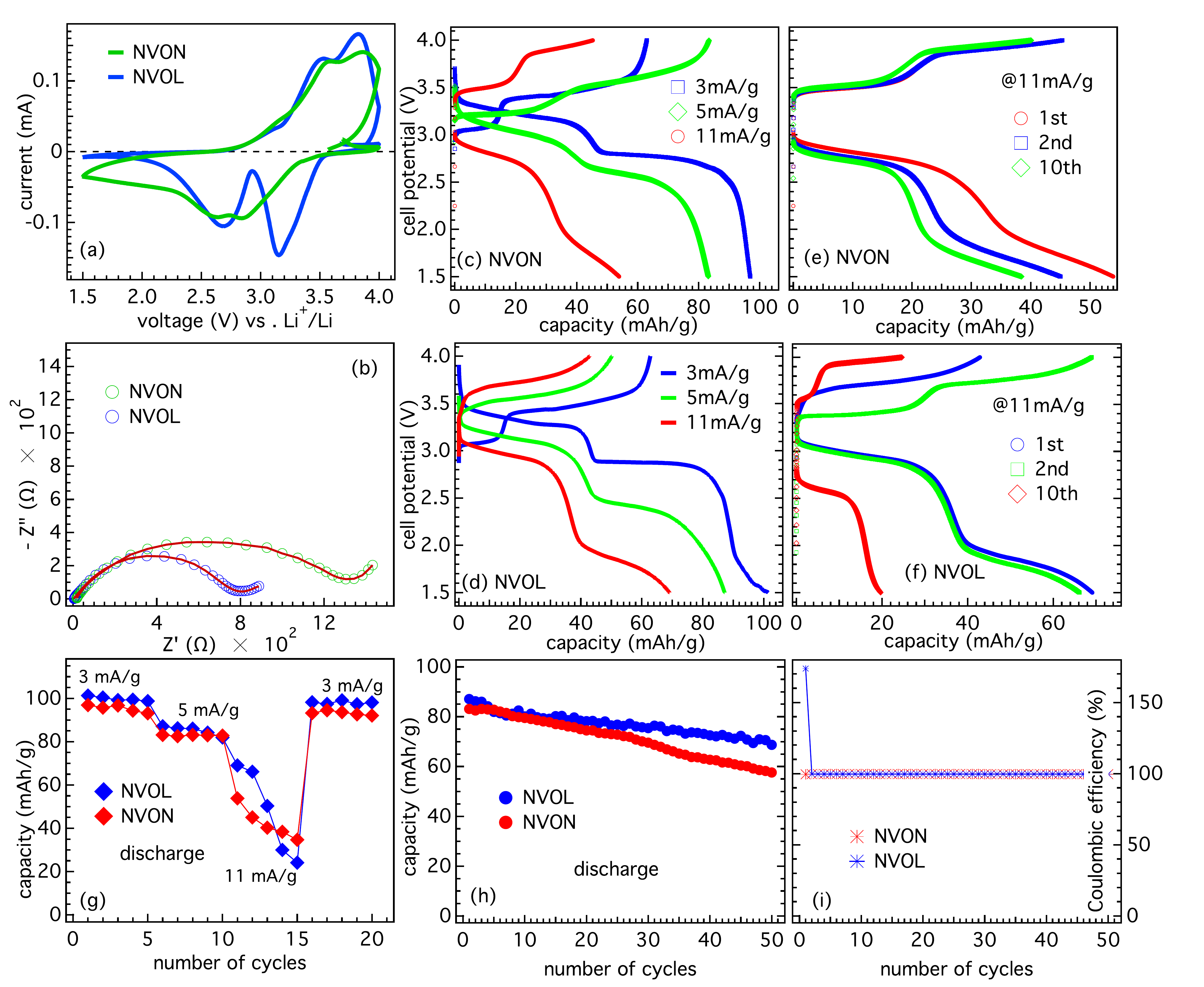}
\caption{ (a) Cyclic voltammetry, and (b) EIS results of the NVO half-cell assembled with NaClO$_4$ (NVON) and LiPF$_6$ (NVOL) electrolyte vs. Li/Li$^+$. Initial charge/discharge comparison at different current densities for (c) NVON (d) NVOL. The charge/discharge comparison of 1st, 2nd and 10th cycles measured at 11~mA g$^{-1}$ for (e) NVON (f) NVOL. (g) The rate performance of the NVOL and NVON half-cells at different current densities. (h) The cycling performance of the half-cells at current density of 5 mA g$^{-1}$ for 50 charge/discharge cycles, and (i) the Columbic efficiency of the half-cells.} 
\label{fig:Electrochemical1}
\end{figure*}

In order to test the feasibility of prepared Na$_{0.66}$V$_4$O$_{10}$ (NVO) as a cathode material for Li-ion batteries, we assembled the half-cells vs. Li/Li$^+$ as an anode and performed the electrochemical measurements with two electrolytes i.e. 1 M NaClO$_4$ and LiPF$_6$ (EC:DMC 1:1 V/V) to find which one is better suited for this purpose. These cells are abbreviated as NVON (for NaClO$_4$) and NVOL (for LiPF$_6$), respectively. The Cyclic voltammograms of the prepared NVO vs. Li/Li$^+$ cells are shown in Figure~4(a) for both the electrolytes. The data was recorded at a scan rate of 0.1 mVs$^{-1}$. Both the half-cells show well defined oxidation/reduction peaks [Figure~4(a)]. However, the intensity of redox peaks in the NVOL half-cell is higher than those of NVON redox peaks, which could be attributed to the better Na-ion diffusion. The redox pairs observed at 3.5/2.7~V, 3.8/3.1~V (for NVOL) and 3.5/2.7~V, 3.9/2.8~V (for NVON) can be due to intercalation of Li-ions in the host NVO structure [Figure~4(a)]. The EIS results of both the half-cells are presented in Figure~4(b), where it can be observed that NVOL half-cell possess lower charge-transfer resistance as compared to the NVON half-cell, which is well consistent with the CV results. The charge-resistance was found to be around 800 and 1300~$\ohm$ for the NVOL and NVON half-cells, respectively [Figure~4(b)]. The CV and EIS results indicate that NVOL half-cell would display the better electrochemical properties as compared to the NVON half-cell.

Figures~4(c, d) show the initial charge/discharge curves for both the half-cells (NVOL and NVON) and the results are in well agreement with CV and EIS results. One slopping voltage plateau around 3.5~V and other flat plateau around 2.7~V can be clearly seen. However, as the current density increases from 3~mA g$^{-1}$ to 11~mAh g$^{-1}$, the voltage plateaus shift to the lower side due to an increase in the polarization and the cell resistance. The specific discharge capacity was found to be around 101, 87 and 69~mAh g$^{-1}$ at the current density 3, 5 and 11~mA g$^{-1}$, respectively for NVOL half-cell. On the other hand, for NVON half-cell the obtained values are around 97, 83 and 55 mAh g$^{-1}$ at the current density 3, 5 and 11~mA g$^{-1}$, respectively. The intercalation and deintercalation of the Na-ion in the host structure can be understood in the following reaction:
During first discharge process, Li-ion moves in to the host NVO electrode:
\begin{equation}
\rm{Na}_{0.66}V_4O_{10} + xLi^+ + xe^- \rightarrow Li_xNa_{0.66}V_4O_{10}
\end{equation}
During the charging cycle, Na$^+$ ions together with Li$^+$ ions will be removed from the host structure:
\begin{equation}
\rm{Li}_xNa_{0.66}V_4O_{10} - xLi^+ - yNa^+ - (x+y)e^- 
\break
\rightarrow \rm{Na}_{(0.66-y)}V_4O_{10}
\end{equation}

Thus obtained Na$_{(0.66-y)}$V$_4$O$_{10}$ will again combined with Li-ion during discharge:
\begin{equation}
\rm{Na}_{(0.66-y)}V_4O_{10} + zLi^+ + ze^- \rightarrow  Li_zNa_{(0.66-y)}V_4O_{10}   
\end{equation}       
Therefore, during repeated charging/discharging, both the Na and Li-ion will contribute in the electrochemical performance of the battery.

In Figures~4(e, f), we show the charging/discharging comparison of the 1$^{st}$, 2$^{nd}$ and 10$^{th}$ cycles for both the half-cells recorded at the current density of 11~mA g$^{-1}$. The NVOL half-cell displays a large capacity fading over consecutive cycles as compared to the NVON cell. The specific discharge capacity of NVOL half-cell decrease from 69~mAh g$^{-1}$ to 20~mAh g$^{-1}$ (~72\% decrease), while in NVON half-cell the decrease is ~31\% (from 53 mAh g$^{-1}$ to 38 mAh g$^{-1}$) in 10$^{th}$ cycle. 
Figures~4(g, h) show the rate and cycling performance of the two half-cells vs. Li/Li$^+$, which show comparable capacities up to 5~mA g$^{-1}$ and for 11~mA g$^{-1}$, the capacity in NVOL cells decays faster as compared to NVON cells, see Figure~4(g). The specific discharge capacity decays ~76\% for NVOL, while for NVON half-cell the capacity decays ~64\%, when the current density changes from 3 ~mA g$^{-1}$ to~11 mA g$^{-1}$. Now we show in Figure~4(h) the cycling performance measured at the current density of 5~mA g$^{-1}$ up to 50 charge/discharge cycles. Both the cells show a gradual decrease in the discharge capacity as capacity decreases from 106 to 69~mAh g$^{-1}$ for the NVOL and 83 to 57~mAh g$^{-1}$ for the NVON half-cell. The Columbic efficiency for both the half-cells is measured at 5~mA g$^{-1}$ current density [Figure~4(i)]. For both the half-cells, we observed 100\% Coulombic efficiency for all the cycles except the first one for NVOL, which is unexpectedly higher 176\%. The possible explanation for such high value in first cycle is that during the discharge there is an extra charge (or electrons) because of the presence of Li in the host structure in the alloy form (Li$_x$Na$_{0.66}$V$_4$O$_{10}$). This extra charge contributes in the capacity during the discharge, while during charging only ${Na}_{0.66}V_4O_{10}$ remains (see equations 4 and 5) and therefore, the charge capacity is lower, which reflect in the calculation of the Coulombic efficiency.

\begin{figure}
\includegraphics[width=3.5in]{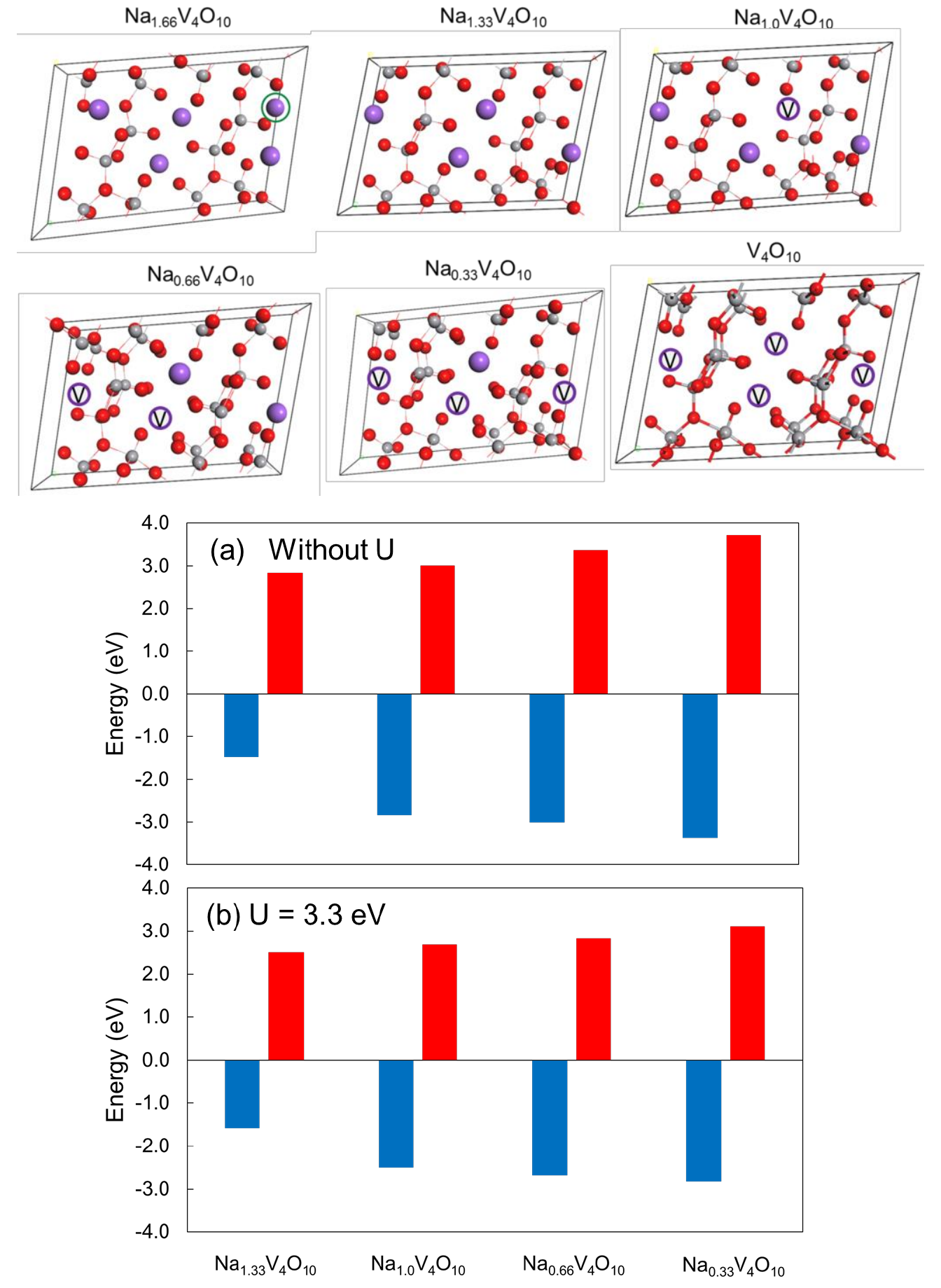}
\caption{DFT optimized geometry of NVO structures with different Na content, showing the position of Na vacancy (violet circle with V) and extra Na (green circle). DFT calculated Na vacancy formation energies (red) on discharging and Na filling energies (blue) on charging in the bulk of the NVO structures with different Na content, (a) without U, and (b) with U$_{eff}=$ 3.3~eV for Vanadium.}
\label{fig:Fig5}
\end{figure}

\begin{figure*}
\includegraphics[width=7.1in]{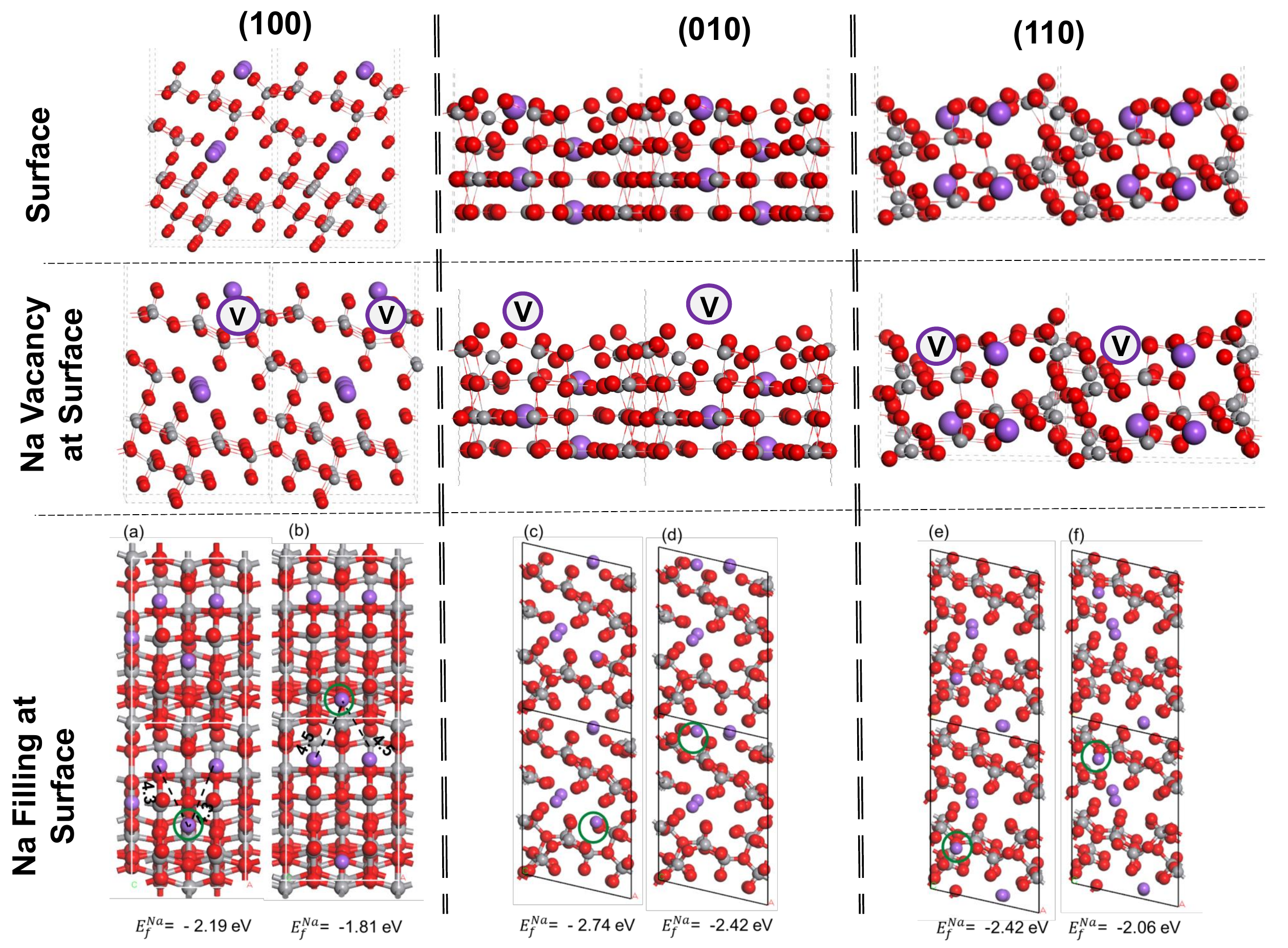}
\caption{DFT optimized geometry of (100), (010) and (110) surfaces of Na$_{0.66}$V$_4$O$_{10}$, showing the position of Na vacancy (violet circle with V) and extra Na (green circle). Lower panels (a--f) show the position extra Na atom during the Na filling (green circle) for different surfaces, energies are indicated at the bottom. The distances between the Na atoms in (a) and (b) are given in Angstrom unit.}
\label{fig:Fig5}
\end{figure*}

\subsection{\noindent ~Density Functional Theory}

The density functional theory (DFT) is used to calculate charging and discharging energy of NVO structures with different Na content. The optimized NVO crystal structures, with different Na content are shown in the upper panel of Figure~5. Here, the structures of Na$_{1.0}$V$_4$O$_{10}$, Na$_{0.66}$V$_4$O$_{10}$, Na$_{0.33}$V$_4$O$_{10}$, and V$_4$O$_{10}$, were obtained by sequential removal of Na atoms, whereas structure of Na$_{1.66}$V$_4$O$_{10}$ was obtained by adding one Na atom to Na$_{1.33}$V$_4$O$_{10}$. The sequential removal of Na atom from Na$_{1.0}$V$_4$O$_{10}$ to Na$_{0.66}$V$_4$O$_{10}$, was done keeping in mind the highest separation of Na vacancy. The other structure obtained by sequential removal of Na atoms are unique as the Na atoms are equivalent. The charging, represents the energy gain by the addition of Na; whereas discharging represents the energy required to remove a Na ion. Thus, charging/discharging process can be defined as:
\begin{equation}
Na_xV_4O_{10} + Na \rightarrow Na_{x+1}V_4O_{10}   \rm \hskip 0.2cm (charging)
\end{equation}
\begin{equation}
Na_xV_4O_{10} \rightarrow Na_{x-1}V_4O_{10} + Na    \rm \hskip 0.2cm (discharging)
\end{equation}
As the number of Na atoms in Na$_x$V$_4$O$_{10}$ decreases from 1.33 to 0.33, a gradual increase in both the energy gain during charging and also the energy required for discharging, is observed. The Na$_{1.33}$V$_4$O$_{10}$ has the lowest Na vacancy formation energy (E$_v^{Na}$ = 2.84~eV), but the energy gain during the charging is low (E$_f^{Na}$ = - 1.48~eV). Hence, Na$_{1.33}$V$_4$O$_{10}$ cannot be a good candidate for the battery operation. Whereas, Na$_{0.33}$V$_4$O$_{10}$ shows a very high energy storage capability due to its high energy gain during the Na addition (E$_f^{Na}$ = - 3.37~eV ), but it also have a very high Na vacancy formation energy (E$_v^{Na}$ = 3.71 eV), hence also not a good candidate for battery application. The Na$_{0.66}$V$_4$O$_{10}$ samples have the moderate values for both the energy gains in Na addition (E$_f^{Na}$ = - 3.01~eV) and the energy required for creating Na vacancy (E$_v^{Na}$ = 3.37~eV), see Figure~5(a) for calculations performed without considering U for Vanadium. These two characteristics directly point towards the potential of Na$_{0.66}$V$_4$O$_{10}$ as a good battery candidate. However, the energy required for creating Na vacancy at the bulk of Na$_{0.66}$V$_4$O$_{10}$ is still very high (E$_v^{Na}$= 3.37~eV, Table~II) and will be difficult at the normal battery operating conditions. 

\begin{table}[h]
  \centering
  \caption{The calculated energy gain due to the addition of Na at the different surfaces.}
  \vskip 0.2 cm
  \label{tab:MH}
   \begin{tabular}{|c|c|c|c|}
  	\hline
   NVO  & Surface & Na vacancy  & Na filling \\
   (Na$_{0.66}$V$_4$O$_{10}$) & energy  & formation & energy \\
  &  (eV/\AA$^2$) & energy (eV) & (eV)\\
    \hline
   Bulk &--- &3.37 &-3.01  \\
    \hline
    (100) &0.031 &2.52 &-2.19  \\
    \hline
    (010)&0.039 &3.36 &-2.74  \\
    \hline
    (110) &0.046 &3.01 &-2.47  \\
    \hline
  \end{tabular}
\end{table}

In order to check the effect of U on the Na vacancy formation energy and Na filling energies were recalculated with U$_{eff}=$ 3.3~eV applied for Vanadium \cite{Jain13}, see Figure~5(b). We observed the similar trend on the Na vacancy formation energy and filling energy of the Na$_x$V$_4$O$_{10}$ sample with U$_{eff}=$ 3.3~eV for vanadium.

The nanostructured morphologies of the cathode materials are in general known to show better electrochemical activities \cite{KangJMC12, QiACI15, HayashiESSL98}, as also shown in our experimental results. It is understood that the nano-structuring opens up the surface diffusion channels, where the formation of the Na vacancy and the diffusion of Na are known to be easy as compared to the bulk. In order to understand the nano-structuring phenomena for our target NVO material Na$_{0.66}$V$_4$O$_{10}$, we studied the surface Na vacancy formation energy and Na filling energies at three different surfaces of NVO, (100), (010) and (110) as shown in Figure~6. The DFT calculated energies for Na vacancy formation and Na filling are listed in Table~II, for all the three surfaces. The surface energy calculation shows that the (100) surface is more stable than the (010) and (110) surface, and hence will cover most of the exposed surface of the nano scale material. The Na vacancy formation energy is calculated to be lowest for the (100) surface (E$_v^{Na}$ = 2.52~eV); which is lower by 0.85 eV, compared to the bulk (E$_v^{Na}$ = 3.37~eV). The (010) surface has the similar Na vacancy formation energy as the bulk; whereas on (110) surface it is a little lower (E$_v^{Na}$= 3.01~eV). The lower values of Na vacancy formation energy at the (100) surface, will enhance the concentration of Na vacancy at the surface; which will result in a higher diffusion and hence better electrochemical performance. 

The Na filling energy also follows the same trends as the Na vacancy formation energy. The energy gain due to the addition of Na at the (100) surface is calculated to be the lowest (E$_f^{Na}$ = - 2.19~eV); whereas the energies are also lower for (010) and (110) surfaces (E$_f^{Na}$ = -2.74~eV and - 2.47~eV, respectively) as compared to the bulk (E$_f^{Na}$ = - 3.01~eV), as given in Table~II. Furthermore, for the calculations of Na vacancy formation energy and Na filling energy, we have tested different possible configurations and the most stable configuration was used to calculate the Na vacancy formation energy and Na filling energy. The two Na atoms present at the (100) surface are equivalent, hence unique geometry was obtained for Na vacancy formation energy at the (100) surface. For the Na atom filling energy at the (100) surface, two configurations were studied as shown in the lower panel of Figure~6 (a, b) along with the filling energy. For the (010) surface there is only one Na atom present at the surface, hence the geometry obtained for Na vacancy formation energy at the (010) surface is unique. Similar to the (100) surface, two different configurations were obtained for the Na atom filling energy. On the other hand, the two Na atom present at the (110) surface are equivalent, hence only one unique geometry was obtained for Na vacancy formation energy. Also, two different configurations were obtained for the Na atom filling energy at the (110) surface, see Figure~6 (e, f).

\section{\noindent ~Conclusions}

In summary, the rod-shaped Na$_{0.66}$V$_4$O$_{10}$ (NVO) nanostructured cathode material was prepared using sol-gel method. The XRD patterns confirm the formation of pure monoclinic phase with the C2/m space group and no impurity or secondary phases were observed. The SEM and TEM studies reveal the rod-shaped agglomerated morphology of the prepared cathode material with size in 50--100~nm range and the length a few $\mu$m. The observed selected area electron diffraction pattern is in good agreement with the XRD pattern. The cyclic voltammetry (CV) results showed the two-phase transition reaction between V$^{5+/4+}$, which is in good agreement with the galvanostatic charge/discharge curves. The electrochemical performance indicates that the NVO cathode exhibits specific discharge capacity of 80 ($\pm$2), 64 ($\pm$2), 56 ($\pm$2), 52 ($\pm$2) and 30 ($\pm$2) mAh g$^{-1}$, measured vs. Na$^+$/Na at the current densities of 3, 5, 11, 20 and 50 mA g$^{-1}$, respectively. The electrochemical performance of Na$_{0.66}$V$_4$O$_{10}$ electrode with Li anode is found promising, but decays faster as compared to the Na-anode. The energetics of Na vacancy formation and filling on discharging and charging, respectively, which could form a potential descriptor of the electrochemical performance. The available surface area for Na transport is likely to increase for nanostructures, which further supports the experimental results.

\section{\noindent ~MATERIALS AND METHODS}

{\bf Experimental details}. The nanostructured Na$_{0.66}$V$_4$O$_{10}$ (NVO) samples were synthesized by sol-gel route using sodium acetate (C$_2$H$_3$NaO$_2$) and ammonium metavanadate (H$_4$NO$_3$V) in 1 to 6.06 molar ratio, from Sigma Aldrich, $\geq$99.0\%. The starting precursors were homogeneously mixed in deionized water with continuous stirring. Citric acid (Anhydrous, C$_6$H$_8$O$_7$, from Fisher Scientific, 99.5\%) was used as a chelating agent in a molar ratio of 1:1 with metal ions. The resulting mixture of citric acid and precursors was heated at 80$^{\rm o}$C with continuous stirring at 400~rpm until a homogenous gel was obtained. Then, the gel was dried overnight at 120$^{\rm o}$C to obtain the powder followed by heat treating in muffle furnace in air at 400$^{\rm o}$C (with 5$\degree$C/minute rate) for 5 hrs. The final powder was stored in a vacuum desiccator for physical and electrochemical characterizations. The thermal decomposition of the resulting gel precursors was examined by thermogravimetric analysis (NETZSCH, TG 209 F3 Tarsus) at a ramp rate of 10$^{\rm o}$C/min in air from RT to 900$^{\rm o}$C. 

The crystallographic structure and phase purity of the prepared samples were determined by room temperature powder x-ray diffraction (from Pananlytical) with CuK$\alpha$ radiation ($\lambda$=1.540 \AA) operating at 40~kV voltage and 40~mA current. The XRD pattern was recorded in 2$\rm \theta$ angular range between 10 and 70$^{\rm o}$ with a step size of 0.02$^{\rm o}$. The XRD data have been analyzed by Rietveld refinement using FullProf package and the background was fitted using linear interpolation between the data points. We study the morphology of the prepared cathode using Zeiss EVO 50 scanning electron microscope (SEM) working at 20~kV in scattered electron mode. A JEOL JEM-1400 Plus microscope coupled with energy dispersive x-ray spectroscopy (EDX) facility has been used for the transmission electron microscopy (TEM) measurements at 120~keV. 

For the electrochemical measurements, cathodes were prepared by mixing prepared active material, PVDF (Polyvinylidene difluoride) as a binder and super P (conductive carbon) in a weight ratio of 80:15:5 in NM2P (N-methyl-2-pyrrolidinone) as a solvent. Then we stirred the resulting mixture for 5~hrs to have homogenous mixing and then coated over Al foil (as a current collector) using doctor blade technique. The coated Al sheet was then dried overnight in vacuum oven at 80$^{\rm o}$C to evaporate the solvent. The sheet was then rolling pressurized and punched into circular electrode with 12~mm diameter. The active mass loading on the electrodes was between 2-3~mg. Before bringing inside the glove box workstation, we dried the electrodes in vacuum at 60$^{\rm o}$C to remove any moisture. Na chips (16~mm diameter) were extracted from Na cubes (Sigma Aldrich, 99.9\%) by cutting, rolling and pressing sodium cubes into thin sheets inside the glove box and used as both counter and reference electrode. For electrochemical measurements, the CR2016 coin half-cells were assembled in an argon-filled glove box (Jacomex, $\leq$0.5 ppm of O$_2$ and H$_2$O level) with the cathode electrode, Na metal, and a glass fiber (Advantec, GB-100R) as the separator. The electrolyte used was 1 M NaClO$_4$ dissolved in ethylene carbonate (EC)/dimethyl carbonate (DMC) in a volume ratio of 1:1. Electro-chemical Impedance Spectroscopy (EIS) using an AC voltage pulse of 5~mV in the frequency range of 100~kHz -- 10~mHz was also performed to measure the electrode resistance. We have also assembled and tested the half-cells vs. Li$^+$/Li as an anode to study the feasibility of the prepared NVO cathode as a material for Li-ion batteries. The assembled cells were tested with both electrolytes like 1 M NaClO$_4$ and LiPF$_6$ (EC:DMC 1:1 V/V) and abbreviated as NVON (for NaClO$_4$)  and NVOL (for LiPF$_6$). All the electrochemical measurements were carried out at room temperature using VMP3 (Biologic) instrument. The cyclic voltmmetry (CV) have been performed in the potential window of 1.5 -- 4.0~V vs. Na/Na$^+$ at the scan rate of 0.05~mVs$^{-1}$. The charging/discharging characteristics were studied in galvanostatic mode at different current densities.     

{\bf Computational Method}. Periodic plane-wave based density functional module implemented in Viena ab-initio Simulation Package (VASP) \cite{Kresse96} is applied to calculate the energy of the cathode material with six different Na$_x$V$_4$O$_{10}$ compositions, where $x=$ 1.66, 1.33, 1, 0.66, 0.33 and 0. The model structure of Na$_{1.33}$V$_4$O$_{10}$ [Na$_2$V$_6$O$_{15}$] composition was obtained from ÒThe Materials ProjectÓ database https://materialsproject.org/ (materials ID. mp-778594) \cite{Jain13}, having monoclinic cyrstal structure with C2/m space group. The lattice parameter of $a=$ 15.7845~\AA, $b=$ 3.6662~\AA, $c=$ 10.6182~\AA, and $\alpha=$ 90$^{\degree}$ and $\beta =$ 103.093$^{\degree}$, matches well with the experimentally obtained cell parameters. The sodium bulk metal geometry is modelled as cubic {\it bcc} bulk with lattice parameter, $a=$ 4.2906~\AA~ and $\alpha=$ 90$^{\degree}$. Geometry optimization is performed on the structures, with energy and force convergence criteria set to 1$\times$10$^{-4}$~eV and 0.05~eV/\AA, respectively. RPBE \cite{HammerPRB99} GGA exchange correlation functional and Ultra-soft pseudopotentials (USSP) \cite{VanderbiltPRB90} are used. Plane wave basis sets are expanded to an energy cut-off value of 396~eV. Monkhorst pack \cite{MonkhorstPRB76} k-points grid of 1$\times$3$\times$1 is utilized for all the calculations on different NVO structures, whereas a 5$\times$5$\times$5 grid is used for Na bulk lattice geometry optimization.
The Na vacancy formation energy (E$_v^{Na}$) of NVO structure Na$_x$V$_4$O$_{10}$ is calculated using the formula below;
\begin{equation}
E_v^{Na}=E[Na_{(x-1)}V_4O_{10}]+E(Na)- E(Na_xV_4O_{10}), 
\end{equation}

where E(Na$_x$V$_4$O$_{10}$), E(Na$_{(x-1)}$V$_4$O$_{10}$) and E(Na) are the energies of NVO structure Na$_x$V$_4$O$_{10}$, Na$_{(x-1)}$V$_4$ O$_{10}$ formed after Na removal and energy of bulk Na, respectively.
Similarly the energy gain from Na addition (E$_f^{Na}$) to NVO structure Na$_x$V$_4$O$_{10}$ can be calculated as;
\begin{equation}
E_f^{Na}=E(Na_{(x+1)}V_4 O_{10})- E(Na_xV_4O_{10})- E(Na)		
\end{equation}

In order to study the nano-structuring effect on the Na vacancy formation and Na filling energies; three different low-index surfaces, (100), (010) and (110) are created for the NVO structure studied experimentally, Na$_{0.66}$V$_4$O$_{10}$. A 2$\times$1 super cell was used for (100) surface, whereas for both the (010) and (110) surfaces, a 1$\times$1 unit cell is used for the surface calculation. For all the three surfaces, a four layer surface slab is used; where two bottom layers are kept fixed to their bulk positions and the top two layers are allowed to relax. A vacuum of 20~\AA~ is used to model the surface slabs. Calculations of Na vacancy formation energy and Na filling energy are performed by removing and adding Na at the corresponding surface, respectively. For the addition of Na to the surface two distinct adsorption sites were studied and the geometry and energy of the most stable adsorption site was reported. Equations 9 and 10 are used to calculate the surface Na vacancy formation energy and Na filling energy, respectively.

\section*{\noindent ~Acknowledgments}

We acknowledge the financial support from IIT Delhi through the FIRP project (IRD no. MI01418). RS thanks IRD, IIT Delhi for postdoctoral fellowship through FIRP project. MC, and RS thank SERB-DST (NPDF, no PDF/2016/003565), and MHRD, respectively, for the fellowship. The authors thank the IIT Delhi for providing central research facilities: XRD, EDX, SEM, and TEM. We also thank the physics department, IIT Delhi for support. RSD acknowledges the financial support from SERB-DST through Early Career Research (ECR) Award (project reference no. ECR/2015/000159) and BRNS through DAE Young Scientist Research Award (project sanction no. 34/20/12/2015/BRNS). The authors thank the HPC facility of IIT Delhi for computational resources.

\end{document}